\documentclass[aps,prl,showpacs,noshowkeys,amsmath,amssymb,amsfonts,reprint]{revtex4-1}
\usepackage{graphicx}
\usepackage{dcolumn}
\usepackage{bm}
\usepackage{color}

\begin{document}
\title{Orientational Dynamics of Fluctuating Dipolar Particles \\
Assembled in a Mesoscopic Colloidal Ribbon}
\author{Helena Massana-Cid$^1$}
\author{Fernando Martinez-Pedrero$^{2}$}
\author{Andrejs Cebers$^3$}
\author{Pietro Tierno$^{1,4,5}$}
\email{ptierno@ub.edu}
\affiliation{
$^1$Departament de F\'isica de la Mat\`eria Condensada, Universitat de Barcelona, 08028 Barcelona, Spain.\\
$^2$Departamento de Qu\'imica F\'isica I, Universidad Complutense de Madrid, Ciudad Universitaria, 28040, Madrid, Spain.\\
$^3$University of Latvia, Faculty of Physics and Mathematics, Zellu 23, LV-1002.\\
$^4$Universitat de Barcelona Institute of Complex Systems (UBICS), Universitat de Barcelona, 08028 Barcelona, Spain\\
$^5$Institut de Nanoci\`encia i Nanotecnologia, IN$^2$UB, Universitat de Barcelona, 08028 Barcelona, Spain.}
\date{\today}
\begin{abstract}
We combine experiments and theory
to investigate the
dynamics and orientational fluctuations of
ferromagnetic microellipsoids that form a ribbon-like
structure due to attractive dipolar forces. 
When assembled in the ribbon, 
the ellipsoids displays 
orientational 
thermal fluctuations 
with an amplitude that can be controlled 
via application of an in-plane magnetic field.
We use video microscopy 
to investigate the orientational dynamics 
in real time/space. 
Theoretical arguments are used to 
derive an analytical expression 
that describes how the distribution of the different angular 
configurations depends on the strength of the applied field.
The experimental data are in good agreement with
the developed
model for all the range of field parameters explored.
Understanding the role of fluctuations in
chains composed of dipolar particles is important not only
from a fundamental point of view,
but it may also help understanding
the stability
of such structures against thermal noise, which is 
relevant
in microfluidics and lab-on-a-chip 
applications.
\end{abstract}
\maketitle
\section{Introduction}
Brownian particles
assembled into a linear chain due to 
anisotropic interactions,
as the ones arising from
dipolar forces,
represent an accessible
and thus appealing model
system to study
the role of noise in
simple polymer-like structures~\cite{Doi88}.
When the linkage between the particles
is not provided by
a strong chemical bond~\cite{Fur98,Gou03,Bis03,Vut11,Dem14},
but results from a weak attractive 
interaction~\cite{Skj83,Tou04},
then the fluctuations of the single 
particles may
significantly influence on the chain dynamics,
producing
torsions, bending or even irreversible breakage.
Investigating the role of thermal noise
in such systems,
and how the 
single particle fluctuations affect the chain dynamics,
is thus necessary to understand the
behaviour and the stability
of the whole structure.

There are different
works that explored the deformations and the dynamics  
of chains composed by
spherical microspheres~\cite{Sil96,Cut01,Hon06,DLi11}.
More recently,
experiments
with magnetic dumbbells~\cite{Zer08},
Janus rods~\cite{Yan13} and
hematite ellipsoids~\cite{Mar16},
have shown the possibility to
realize and manipulate elongated
structures composed by anisotropic
colloids via external fields.
In contrast to spherical colloids,
anisotropic particles such as ellipsoids
introduce an additional
rotational degree of
freedom that complicates their dynamics, 
giving rise to a
richer physical behavior.
However, the role of thermal fluctuations in the
orientation of the anisotropic elements 
when assembled into linear chains
has not been addressed yet.

In this article, we study the 
dynamics of ferromagnetic microellipsoids
around the direction determined by an 
external field, both alone and when assembled
into a ribbon. 
In the latter case, the orientational fluctuations are 
described by formulating 
a theoretical framework 
based on a modified version of 
the worm-like
model~\cite{Doi88}.
The model allows capturing the
fundamental physics of the 
process,
and deriving an analytic expression 
for the distribution of the 
particle
orientations within the ribbon.

\section{Experimental system and procedures}
The anisotropic ferromagnetic
ellipsoids are synthesized
following a well established procedure
developed by Sugimoto and co-workers~\cite{Sug93},
and described in details
in several previous works~\cite{Ger08,Lee09,Pal13,Mar162}.
With this method we obtain monodisperse
hematite particles 
with prolate shape and
a major (minor) axis equal to $a=1.80\pm 0.11 \, \mu m$
($b=1.31 \pm 0.12\, \mu m$ resp.).
From the analysis of scanning electron microscopy images, we measure a polydispersity index equal to $\sigma_a= 0.022$ and $\sigma_b= 0.023$, for the long and short axis of the synthesized particles respectively.  
After synthesis, the particles have
a small permanent moment $m \simeq 2 \times  10^{-16} \, Am^2$,
perpendicular to their
long axis, as depicted in Fig.1(a).
The peculiar orientation of the permanent moments, 
as compared to other 
anisotropic magnetic particles~\cite{TiePCCP},
is caused by the magnetic
structure of hematite,
which crystallizes in the corundum form~\cite{Shu51}.
The value of $m$ was obtained by measuring the reorientational motion of individual hematite particles subjected to a static magnetic field~\cite{Mar16}.
The permanent moment of the particles 
allows us to estimate a 
dipolar coupling constant $\lambda=\mu_0 m^2/(4 \pi k_B T b^3)=0.44$
and a Langevin parameter 
equal to $\Lambda=\mu_0 mH/(k_B T)=93.1$, for an external field 
$H=1500 \rm{A m^{-1}}$~\cite{Ros86}.
Here $\mu_0=4 \pi 10^{-7} \rm{H m^{-1}}$ is the magnetic permeability, 
$T \sim 293 \rm{K}$ the room temperature and $k_B$ the Boltzmann constant.  
Moreover, from SQUID (Superconducting Quantum Interference Devices)
measurements (data not shown here),
we find that the particle permanent moment is one order of magnitude 
larger than the induced moment in all the range of explored field 
strengths~\cite{Fersmall}. 

After synthesis, the particles are
dispersed in highly deionized water
(purified using a Milli-Q system, Millipore), stabilized with a surfactant
by adding $0.11 \rm{g}$ of solution sodium
dodecyl sulfate
for $80$ ml of water,
and finally the pH of the solution is adjusted to $9.5$
by adding Tetramethylammonium Hydroxide.
These procedures are used to create a protective
steric layer around the particles 
that avoids the irreversible sticking 
due to attractive Van der Waals interactions.
The particles sediment
close to a glass plate, where they
remain quasi two-dimensionally confined due to 
the balance between gravity and
the electrostatic repulsion with the
glass surface.
The particle dynamics are visualized
using an optical microscope (Eclipse Ni, Nikon),
and their positions and orientations are recorded
with a CCD camera (Scout scA640-74f, Basler) working at $50$ fps.
The anisotropic shape of the particles allows for monitoring the 
instantaneous direction of their permanent moment. 
The external magnetic
field is generated 
with a
custom-made coil
system connected to a direct current 
power supply (EL
302RT, TTi).

%
\begin{figure}[!tb]
\begin{center}
\includegraphics[width=0.95\columnwidth,keepaspectratio]{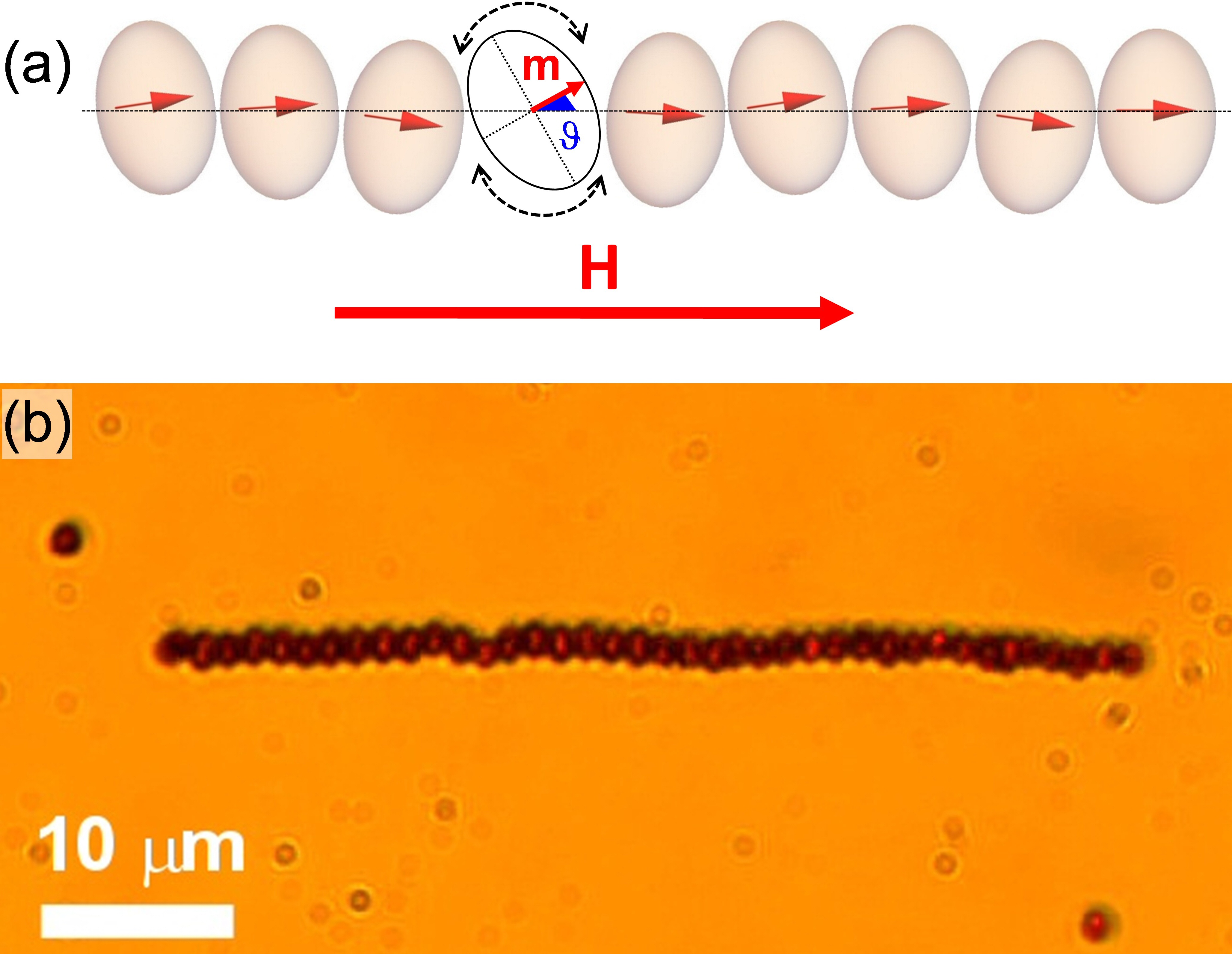}
\caption{(a) Schematic showing a chain
of ferromagnetic ellipsoids
subjected to an external
magnetic field $\bm{H}$,
being $\vartheta$ the angle
between the particle moment and the $x$-axis, 
that coincides with the direction of 
the applied field. 
Each particle
displays orientational thermal fluctuations 
around this axis.
(b) Optical microscope image of
a chain composed by $39$ ellipsoids under
a constant field $H=1600 \, A m^{-1}$, applied along the
same direction as shown 
in the schematic at the top.}
\label{fig_1}
\end{center}
\end{figure}
%

\section{Single particle dynamics}
Before analyzing the orientational fluctuations in the
ribbon, we have
first characterized the thermal motion
of a single ellipsoid in absence and in presence 
of an external field.
The dynamics of an individual ellipsoid
in water has been
treated in different works~\cite{Per34,Han06,Gri07,Gue10,Fan17},
and thus here will be only briefly described.
In the absence of a magnetic field,
the anisotropic shape
of the particle causes a non trivial coupling between
its rotational and translational motion.
In particular, the particle translational diffusion is anisotropic at short times, $t \sim 0s$, being the time interval $t$ defined with respect to the initial measured position of the ellipsoid,
with two different
diffusion coefficients, $D_{\parallel}$ and $D_{\perp}$,
characterizing the translational dynamics
parallel and perpendicular to the particle
long axis, respectively.
This behavior arises from the fact that 
the ellipsoid has
two different friction coefficients,  $\gamma_{\parallel}$ 
and $\gamma_{\perp}$, 
along the
direction
parallel and 
perpendicular to its long axis resp.
Since $\gamma_{\perp} > \gamma_{\parallel}$
and $D=k_B T/\gamma$ one obtain 
thus $D_{\perp} < D_{\parallel}$.
In contrast after long time the 
diffusion becomes isotropic, and the 
two diffusion coefficients coincide 
$D_{\perp} = D_{\parallel}$.
The crossover time
between both behaviors
is given by the
rotational diffusion time,
$\tau_{\vartheta}=1/(2D_{\vartheta})$,
being $D_{\vartheta}$ the
rotational
diffusion coefficient.
Thus, for
time $t < \tau_{\vartheta}$ 
one has that $D_{\parallel}\neq D_{\perp}$,
since the rotational motion still needs to
become important in the particle dynamics.
When $t > \tau_{\vartheta}$ 
the rotational movement erases the directional memory 
of the particle,
and the two diffusion coefficient 
approach a common value.
In order to determine the characteristic time $\tau_{\vartheta}$
for our particles, we
record the motion
of several
individual ellipsoids ($i=1...N$),
and extract the particle positions and orientations, $(x_i(t),y_i(t),\vartheta_i(t))$,
being $\vartheta_i(t)$ the angle between the $x-$axis and the particle 
minor axis.
We then determine the corresponding
diffusion coefficients from the mean square displacement (MSD).
For the angular variable, 
the MSD can be written as, 
\begin{equation}
D_\vartheta (t)=\frac{1}{2Nt}\langle \sum^N_{n=1}[\vartheta_{i}(t)-\vartheta_{i}(0)]^2\rangle \, \, .
\end{equation}
From these data, we measure $D_{\vartheta}=0.08 \, \rm{rad^2 s^{-1}}$
that corresponds to
$\tau_{\vartheta}=6.2 \, \rm{s}$.
We note that this value is larger than
what found for
non-magnetic ellipsoids ($\tau_{\vartheta}=3.1 \, \rm{s}$~\cite{Han06}),
probably due to the different aspect ratio 
and the higher density of the hematite ellipsoids
that could hinder diffusion
by forcing the particle to stay closer to the
substrate,
thus increasing 
the values of the translational and the rotational friction coefficients.
For time $t > \tau_{\vartheta}$, we measure
a common diffusion coefficient, $D_{\parallel}=D_{\perp}=0.074\,\rm{\mu m^2 s^{-1}}$.

An external magnetic field
can be used to impede the rotational motion,
separating the diffusion coefficients
along the directions parallel and perpendicular to the particle's long axis.
In this situation, the particle aligns with its short axis (long axis) parallel (perpendicular) 
to the field direction. 
Thus, there is no transition towards the isotropic
diffusion, and the hematite ellipsoids present an 
anisotropic motion with two 
different diffusion coefficients, 
even at long times.
In particular, for a field amplitude
of $H= 1500 \, \rm{A m^{-1}}$,
we find that the diffusion is totally anisotropic, with 
$D_{\parallel}=0.112 \, \rm{\mu m^2 s^{-1}}$ and $D_{\perp}=0.037 \, \rm{\mu m^2 s^{-1}}$.
We next analyze, first theoretically 
and later via experiments, the dynamics of a chain 
of interacting hematite ellipsoids. 

\section{Theoretical model}
We consider
a chain of permanent dipoles
aligned along 
the direction imposed by a constant field
$H$ and
subjected to thermal fluctuations
that disorder the
particle orientations
by increasing the angle $\vartheta$, Fig.1(a).
We describe this situation by
using a modified version of the
stretched
worm-like
model~\cite{Doi88,Fix73,Mar95}. In our model the energy of the chain is given by:
\begin{equation}
E=\frac{K_{b}}{2}\int^{L}_{0}\Bigl(\frac{d\vartheta}{dl}\Bigr)^{2}dl-MH\int^{L}_{0}\cos{(\vartheta)}dl  \, \, ,
\label{Eq:1}
\end{equation}
where $K_b$ the bending constant that takes
into account for the dipolar interactions
between the ellipsoids
and can be written as $K_b=m^2/2b^2$,
being $b$ the ellipsoid minor axis.
Further,  $M$ is the particle magnetization
per unit length
and $L$ is the length of the chain.
Our analysis of the chain thermal fluctuations  
is based on the relation between the propagator
of the orientation angles of the dipoles and the chain free energy.
The orientational distribution function of the system can be written as
\begin{equation}
P(\vartheta,l)=\int^{2\pi}_{0} G(\vartheta,l|\vartheta',l')P(\vartheta',l')d\vartheta'   \, \, ,
\label{Eq:2}
\end{equation}
where the function $G$ in the limit $l-l'\rightarrow 0$
is given by the Boltzmann distribution:
\begin{equation}
G(\vartheta,l+\Delta l|\vartheta',l)=\sqrt{\frac{K_{b}}{2\pi k_{B}T\Delta l}}e^{\left(-\frac{K_{b}(\vartheta-\vartheta')^2}{2k_{B}T\Delta l}+\frac{MH\cos{(\vartheta)}\Delta l}{k_{B}T}\right)}
\label{Eq:3}
\end{equation}
In the limit $l\rightarrow l'$, we may derive the
following differential equation for
$P$~\cite{Mar95}:
\begin{equation}
\frac{\partial P}{\partial l}=\frac{1}{2l_{p}}\frac{\partial^{2}P}{\partial \vartheta^{2}}+ \frac{MH\cos{(\vartheta)}}{k_{B}T}P=\hat{H}P  \, \, ,
\label{Eq:4}
\end{equation}
where $l_{p}=K_{b}/k_{B}T$ is the persistence length
of the chain.
As a result, $G$ may be expressed through the eigenfunctions $\psi_{k}$ and
the eigenvalues ${\lambda_{k}}$ of the operator $\hat{H}$ as follows:
\begin{equation}
G=\sum_{k}\exp{(\lambda_{k}(l-l'))}\psi_{k}(\vartheta)\psi_{k}(\vartheta')   \, \, .
\label{Eq:5}
\end{equation}
For infinitely long chains, $L\rightarrow\infty$ ($l=L;l'=0$), only the largest eigenvalue
$\lambda_{1}$ will contribute in Eq.(\ref{Eq:5}), and:
\begin{equation}
G(\vartheta,L|\vartheta',0)=\exp{(\lambda_{1}L)}\psi_{1}(\vartheta)\psi_{1}(\vartheta')   \, \, .
\label{Eq:6}
\end{equation}
Since,
\begin{widetext}
\begin{equation}
\label{Eq:7}
G(\vartheta,L|\vartheta',0)=\int ...\int G(\vartheta,\vartheta_{n-1}|\Delta l) ...G(\vartheta_{1},\vartheta'|\Delta l))d\vartheta_{1} ...d\vartheta_{n-1}
\end{equation}
\end{widetext}
we thus obtain:
\begin{equation}
\int G(\vartheta,L|\vartheta',0)d\vartheta d\vartheta'=Z       \, \, ,
\label{Eq:8}
\end{equation}
where $Z$ is a statistical sum done over 
all the ellipsoids that form the magnetic 
chain. Thus, the total free energy can be expressed
as:
\begin{equation}
F=-k_{B}T\ln{Z}=-k_{B}T\lambda_{1}L   \, \, .
\label{Eq:9}
\end{equation}
For large values of $L$, the orientation distribution function 
can be determined by the eigenfunction 
of $\hat{H}$ corresponding to the largest eigenvalue $\lambda_{1}$:
\begin{equation}
P(\vartheta)\sim \psi_{1}(\vartheta)     \, \, .
\label{Eq:10}
\end{equation}
Now we present the results for the two limiting 
regimes characterized by the amplitude of the applied field. 

\section{Large applied fields}
Let us consider a magnetic chain with length 
$L$ under periodic conditions, $\vartheta(l+L)=\vartheta(l)$.
We can write the angle $\vartheta$ as:
\begin{equation}
\vartheta=\sum_{k}\vartheta_{k}\exp{(ikl)};~k=\frac{2\pi n}{L}   \, \, .
\label{Eq:11}
\end{equation}
For small thermal fluctuations,
the energy of the system may be expressed as:
\begin{equation}
E=\frac{L}{2}\sum_{k}(K_{b}k^{2}+MH)|\vartheta_{k}|^{2}
\label{Eq:12}
\end{equation}
According to the  
Boltzmann principle, 
we can write the probability density
of the  amplitude of the fluctuations
as:
\begin{equation}
\varrho\approx\exp{(-L\sum_{k>0}(K_{b}k^{2}+MH)|\vartheta_{k}|^{2}/k_{B}T)}  \, \, ,
\label{Eq:13}
\end{equation}
and the
spectral amplitude as
\begin{equation}
\langle|\vartheta_{k}|^{2}\rangle=\frac{k_{B}T}{L(K_{b}k^{2}+MH)} \, \, ,
\label{Eq:14}
\end{equation}
and
\begin{equation}
\langle\int^{L}_{0}\cos{(\vartheta)}dl\rangle=L-L\sum_{k>0}|\vartheta_{k}|^{2}    \, \, .
\label{Eq:15}
\end{equation}
The sum in Eq. (\ref{Eq:15}) can be calculated 
upon direct integration,
\begin{equation}
\langle\int^{L}_{0}\cos{(\vartheta)}dl\rangle=L\Bigl(1-\frac{1}{4}\frac{1}{\sqrt{MHK_{b}/(k_{B}T)^{2}}}\Bigl) \, \, .
\label{Eq:16}
\end{equation}
At equilibrium, the variation of the chain free energy
with the magnetic field may be expressed
as:
\begin{equation}
\frac{\partial F}{\partial H}=-M\langle\int^{L}_{0}\cos{(\vartheta)}dl\rangle    \, \, ,
\label{Eq:17}
\end{equation}
where the right member in Eq.~\ref{Eq:17} is the total magnetic moment of the chain. 
From Eqs. (10),(17) and (18) we derive
the following
relationship:
\begin{equation}
k_{B}TL\frac{\partial \lambda_{1}}{\partial H}=ML\Bigl(1-\frac{1}{2\sqrt{2}\sqrt{\xi}}\Bigr)   \, \, ,
\label{Eq:18}
\end{equation}
which can be used to
test our variational approach,
being here $\xi=MH2l_{p}/k_{B}T$
the dimensionless magnetic energy.
By defining the dimensionless variable 
$\tilde{\lambda}_{1}=\lambda_{1}2l_{p}$, we have:
\begin{equation}
\frac{\partial \tilde{\lambda}_{1}}{\partial\xi}=1-\frac{1}{2\sqrt{2}\sqrt{\xi}}     \, \, ,
\label{Eq:19}
\end{equation}
\begin{figure}[!tb]
\begin{center}
\includegraphics[width=0.9\columnwidth,keepaspectratio]{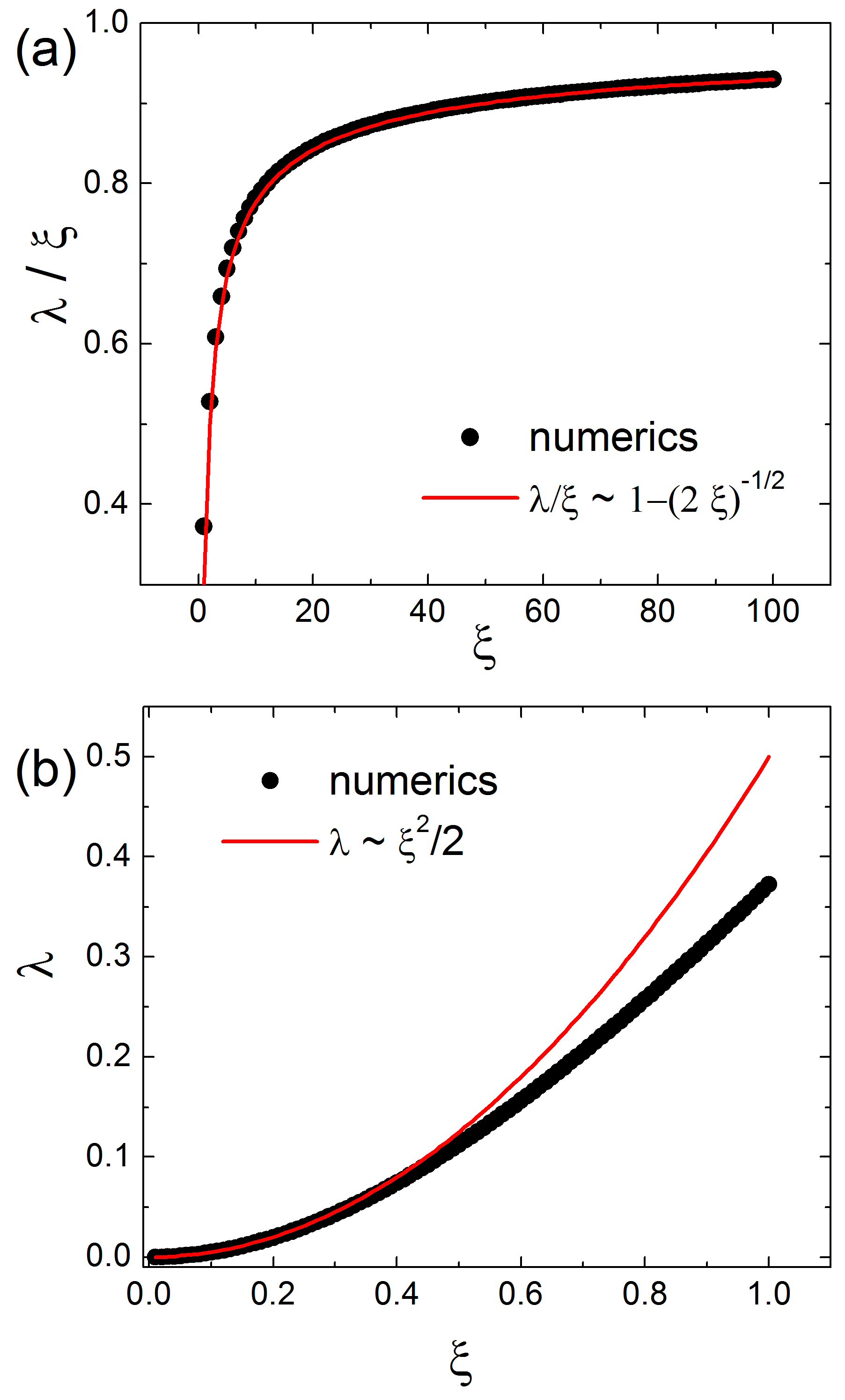}
\end{center}
\caption{(a) Ratio $\lambda/\xi$
versus $\xi$,
being $\lambda$
the eigenvalues
corresponding to the maximum of the 
functional $W$.
The continuous red line 
is a fit corresponding to  Eq.~\ref{Eq:20},
where the
numerical data are obtained for large values 
of $\xi$.
(b) Eigenvalues $\lambda$ 
calculated for small values of $\xi\in[0...1]$.
The continuous red line 
corresponds to Eq.~\ref{Eq:26} 
in the text.}
\label{fig2}
\end{figure}
or
\begin{equation}
\tilde{\lambda}_{1}\simeq \xi-\frac{1}{\sqrt{2}}\sqrt{\xi}   \, \, .
\label{Eq:20}
\end{equation}
\section{Small applied fields}
In the limit of small applied fields,
and according to the perturbation theory, 
the magnetization is given by:
\begin{eqnarray}
\langle\int^{L}_{0}\cos{(\vartheta)}dl\rangle= \nonumber \\
Z^{-1}\int D(\vartheta)\int^{L}_{0}\cos{(\vartheta(l))}dl\exp{(-E/k_{B}T)}\cong\\ \nonumber \frac{MH}{k_{B}T}\langle\int^{L}_{0}\cos{(\vartheta(l))}dl\int^{L}_{0}\cos{(\vartheta(l')}dl'\rangle    \, \, .
\label{Eq:21}
\end{eqnarray}
From Eq.(\ref{Eq:4}) it follows that in the absence of a magnetic field:
\begin{equation}
\langle cos{(\vartheta(l))}\cos{(\vartheta(l'))}\rangle=\frac{1}{2}\exp{(-|l-l'|/(2l_{p}))}   \, \, .
\label{Eq:22}
\end{equation}
Integrating the previous equation one obtain:
\begin{equation}
\int^{L}_{0}\int^{L}_{0}\frac{1}{2}\exp{(-|l-l'|/(2l_{p}))}dldl'\simeq2l_{p}L \, \, .
\label{Eq:23}
\end{equation}
From Eqs. (22), (23) and (24), the magnetic moment is given by:
\begin{equation}
M\langle\int^{L}_{0}\cos{(\vartheta)}dl\rangle=M\frac{MH}{k_{B}T}2l_{p}L  \, \, ,
\label{Eq:24}
\end{equation}
and from Eq.(\ref{Eq:17}):
\begin{equation}
k_{B}T\frac{\partial \lambda_{1}}{\partial H}=M\frac{MH}{k_{B}T}2l_{p} \, \, .
\label{Eq:25}
\end{equation}
Finally, we arrive to another relationship that is  
valid for large thermal fluctuations,
\begin{equation}
\frac{\partial\tilde{\lambda}_{1}}{\partial \xi}=\xi;~\tilde{\lambda}_{1}=\frac{1}{2}\xi^{2} \, \, .
\label{Eq:26}
\end{equation}
and its validity can be demonstrated by 
using a variational method, as described in the following section.
\section{Variational solution}
By defining the dimensionless
length $l/2l_{p}$, Eq.(\ref{Eq:4}) follows from the variation of the functional:
\begin{equation}
W=\int^{2\pi}_{0}\Bigl(-\Bigl(\frac{\partial P}{\partial \vartheta}\Bigr)^{2}+\xi\cos{(\vartheta)}P^{2}\Bigr)d\vartheta/\int^{2\pi}_{0}P^{2}d\vartheta  \, \, .
\label{Eq:27}
\end{equation}
From this equation we have:
\begin{equation}
\frac{d^{2}P}{d\vartheta^{2}}+\xi\cos{(\vartheta)}P=WP   \, \, ,
\label{Eq:28}
\end{equation}
and the largest eigenvalue of the operator $\hat{H}$ corresponds to the maximal value of the
functional $W$. To find this value, we use the following trial function
for the probability density distribution:
\begin{equation}
P=\frac{\exp{(\alpha\cos{(\vartheta)})}}{\sqrt{2\pi I_{0}(2\alpha)}}  \, \, ,
\label{Eq:29}
\end{equation}
where $I_{n}(x)$, $(n=0,1,2,...)$, are the modified Bessel functions
of the first kind.
This function is normalized as, $\int^{2\pi}_{0}P^{2}d\vartheta=1$,
and it is proportional to $\exp{(\alpha\cos{\vartheta})}$.
Using this expression, the functional in Eq.~\ref{Eq:27} can be written as:
\begin{equation}
W=\frac{1}{I_{0}(2\alpha)}\Bigl(-\frac{1}{2}F_{01}(2,\alpha^{2})+\xi I_{1}(2\alpha)\Bigr)   \, \, ,
\label{Eq:30}
\end{equation}
where $F_{01}(2,z)$ is the confluent hypergeometric function.
To justify the selection of the trial function, we
calculated
the values of the parameter $\alpha$
that maximize the functional in Eq.~\ref{Eq:30}  
for both small ($\xi\in [0;0.1]$)
and large ($\xi\in [10;100]$)
values of $\xi$.
The data resulting from these 
calculations are 
plotted versus $\xi$ in Figs.2(a,b).
From these images follow
indeed the numerical data 
are in excellent agreement with
the theoretical expressions 
derived small (Eq.~\ref{Eq:20}) and large (Eq.~\ref{Eq:26}) thermal fluctuations.

\section{Fluctuations in the colloidal ribbon}
In order to validate our model,
we perform different experiments
by measuring the
average angle $\vartheta$ as a function of the applied field.
Thermally induced torsions, which cause the rotation of the ellipsoids 
around the main axis of the chain are 
hindered by the presence of the glass surface, and almost all the 
ellipsoids rest on the surface with their long axis parallel to the horizontal plane. 
In this study we have also neglected the weak 
out-of-plane fluctuations of the magnetic moments. 
Further, we find that
during the measurements  
the magnetic ribbons
are sensitive to  
the earth's magnetic field, $H_{e}\sim 40 \, A m^{-1}$,
and thus 
they orient 
along  
$H_{e}$ even in absence of any extra applied field.
In order to cancel $H_{e}$,
we balance the latter by applying a small DC field in the opposite direction.
Under this condition, the particles in the chain do not present
any preferential orientation, as shown by the empty circles in Fig.3, 
and the ribbons form rings~\cite{Wen99,Yan13,Mar162} or break due to thermal fluctuations.

The distribution of the angles
$P_1(\vartheta)$ are shown in Fig.3,
where the 
experimental data are scattered points, 
while the continuous lines are fit 
to the theoretical model.
Due to the geometry of our 
experimental 
system, we have assumed that the distribution is symmetric 
around the $x-$axis, $P_1(\vartheta)=P_1(-\vartheta)$,
and that $P_1(\vartheta>\pi/2)\sim 0$.
Thus, we normalize the experimental data as:
$\int^{\pi/2}_0 P_1(\vartheta)d\vartheta=1 $,
and use the expression $P_1(\vartheta)=\exp{(\tilde{\alpha}\cos{(\vartheta)})}/(\pi I_{0}(\tilde{\alpha)})$. 
We note that this distribution reduces to 
\begin{figure}[!tb]
\begin{center}
\includegraphics[width=\columnwidth,keepaspectratio]{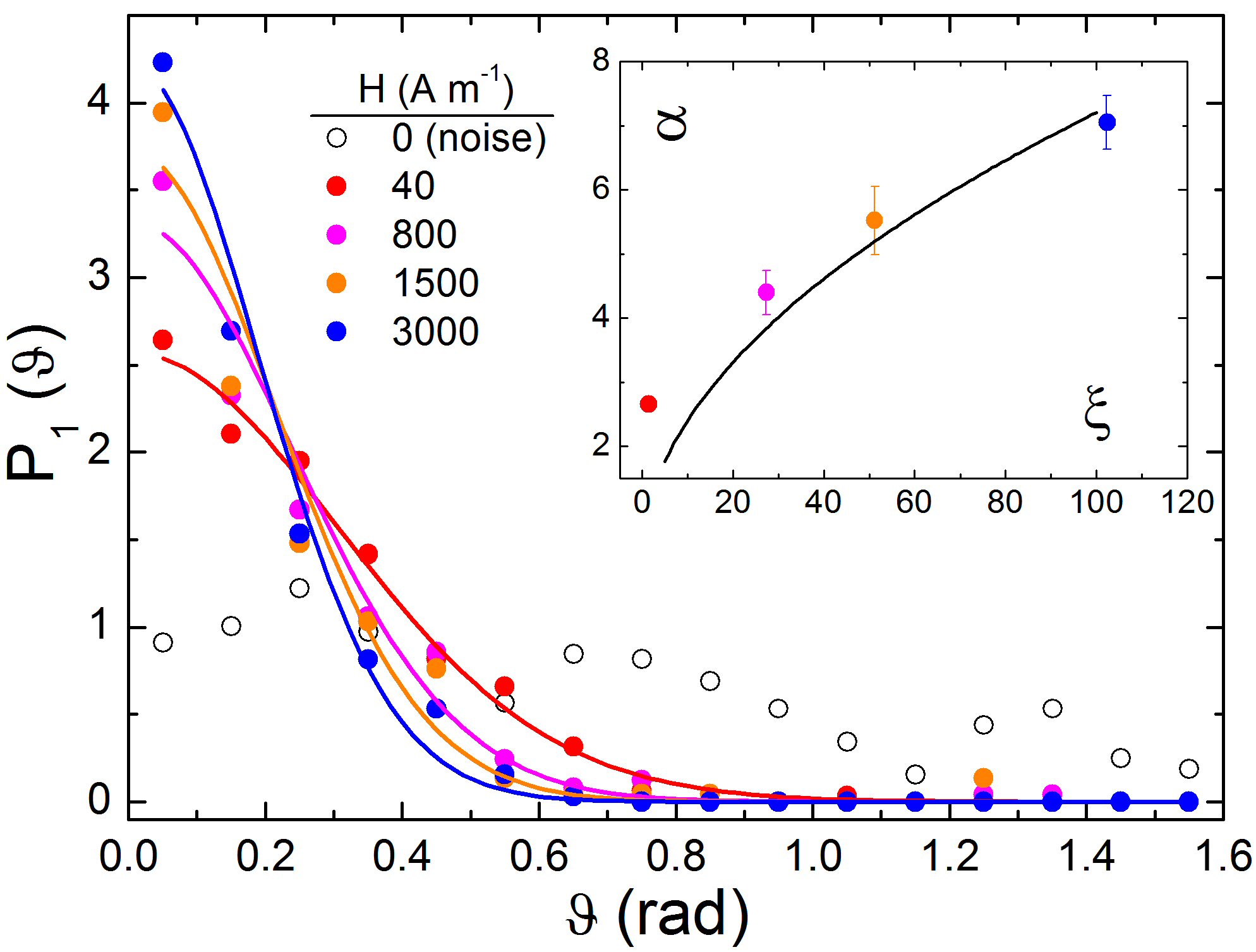}
\end{center}
\caption{Distribution of the particle orientation, $P_1(\vartheta)$,
for different amplitudes of the applied field $H$.
Symbols denote experimental data,
continuous lines
starting from $H=40 \, A m^{-1}$
are theoretical curves calculated following
the equation of $P_1(\vartheta)$ in the text.
The obtained values of the fitted
dimensionless parameter $\alpha=\tilde{\alpha}/2$ for all the curves
are shown as a function of $\xi=mHl_{p}/bk_{B}T$
in the top inset. Here we use $l_p=0.3 \, {\rm \mu m}$
and $m= 2.3 \cdot 10^{-16} \, {\rm A m^2}$.
The continuous line results from
the functional $W$ in Eq.~\ref{Eq:30}.}
\label{fig3}
\end{figure}
Eq.~\ref{Eq:29} of the model $P_1(\vartheta)=P(\vartheta)$ for $\alpha=\tilde{\alpha}/2$
and large values of the parameter $\alpha$.
The application of an external field
reduces the amplitude of the orientational fluctuations,
centering the angular distribution
about $\vartheta=0 \, \rm{rad}$.
The half width of the distribution decreases
with the field, in agreement with 
the prediction of Eq.~\ref{Eq:29}.
The good agreement between the experimental
data and the theoretical expression is also proved
by the small inset in Fig.3.
Here we show
the fitting parameters $\alpha=\tilde{\alpha}/2$
obtained from the main graph
as a function of the normalized magnetic energy 
$\xi$. To determine $\xi$ we use 
as persistence length of the
chain the value $l_p\sim 0.3 \, {\rm \mu m}$,
that was previously determined in a set of independent experiments~\cite{Mar162}.
The scattered data are in excellent agreement with
the continuous line that was independently obtained
from the numerical calculation of the maximum of 
the functional $W$ in Eq.~\ref{Eq:30}.

\section{Conclusions}
We have combined
experiments and theory
to analyze the orientational fluctuations  
in a
colloidal chain
made by ferromagnetic hematite ellipsoids.
Our model quantitatively captures the
physics of the colloidal system, showing
a good agreement with the experimental data.
In absence of external field,
the chain shows large fluctuations
of the individual
elements along the whole range of angles $[0,2\pi]$, 
and can easily break.
When an external field is applied,
the amplitude of the fluctuations decreases, 
becoming  
limited to a narrow
range of angles.

The assembly of magnetic colloids into linear
chains is an
appealing research subject
with both fundamental~\cite{But03}
and technological applications~\cite{Ber96}.
In the first case,
magnetic chains influenced by
thermal fluctuations have
shown a variety of
interesting phenomena,
including
diffusion limited aggregation~\cite{Cer04,Erb09},
multi-scale kinetics~\cite{Swa12}
and subdiffusive dynamics~\cite{Jor11}.
On the application side,
magnetic chains have been used in the past as
micromechanical sensors~\cite{Gou03,Drey09},
to measure the growth of actin filaments~\cite{Bra11},
the rheological properties of the dispersing medium~\cite{Aue06,Hua16},
or even to realize bio-mimetic structures such as
actuated magnetic propellers~\cite{Dre05,Mar115}
and artificial cilia~\cite{Vil10,Coq11}.
Further, our mesoscopic colloidal
system may be also used
as a simplified model for
magnetic polymer beads,
that present exciting applications
in targeted drug delivery 
and in oil industry~\cite{Phi11}.

\begin{acknowledgments}
H.M.C., F.M.P. and P.T. acknowledge support from the
ERC starting Grant "DynaMO".
P.T. acknowledges support 
from MINECO (FIS2016-78507-C2-2-P) and DURSI (2014SGR878).
A. C. acknowledges support from National Research Programme
No. 2014.10-4/VPP-3/21
and from the
M-ERA-NET project "Metrology at the Nanoscale with Diamonds" (MyND).
F.M.P. acknowledges support from MINECO (RYC-2015-18495).
\end{acknowledgments}

\bibliography{biblio}
\end{document}